\documentclass[aip,rsi,reprint]{revtex4-1}

\usepackage{graphicx} 
\usepackage{amsmath}  
\usepackage{comment}  

\begin{document}

\title{Design and experimental validation of a compact collimated Knudsen source}

\author{Steinar H W Wouters}
\author{Gijs ten Haaf}
\author{Peter H A Mutsaers}
\author{Edgar J D Vredenbregt}
 \email{e.j.d.vredenbregt@tue.nl}
\affiliation{Department of Applied Physics, Eindhoven University of Technology, P.O. Box 513, 5600 MB Eindhoven, the Netherlands}

\date{\today}

\begin{abstract}
In this paper we discuss the design and performance of a collimated Knudsen source which has the benefit of a simple design over recirculating sources. Measurements of the flux, transverse velocity distribution and brightness at different temperatures were conducted to evaluate the performance. The scaling of the flux and brightness with the source temperature follow the theoretical predictions. The transverse velocity distribution in the transparent operation regime also agrees with the simulated data. The source was found able to produce a flux of $10^{14}$~s$^{-1}$ at a temperature of 433 K. Furthermore the transverse reduced brightness of an ion beam with equal properties as the atomic beam reads $1.7 \times 10^2$\,A/(m${}^2$\,sr\,eV) which is sufficient for our goal: the creation of an ultra-cold ion beam by ionization of a laser-cooled and compressed atomic rubidium beam.
\end{abstract}


\maketitle 

\section{Introduction}
Collimated sources for atomic species are used in many physics experiments such as laser cooling and compression, spectroscopy and beam collisions. Depending on the application, properties such as total flux, brightness, lifetime and simplicity are important. This paper discusses the design and evaluates the performance of a collimated Knudsen cell that will be used in the atomic beam laser-cooled ion source (ABLIS) \cite{Wouters2014}. Although others have built such sources \cite{Lin2009,Bell2010}, modifications of the design were made to make the device more compact. This paper also presents measurements of the flux, transverse velocity distribution and brightness of the source.
 
The simplest form of a thermal source is the Knudsen cell \cite{ScolesBook}. This device consists of a container in which material is heated to create a vapour pressure. Atoms in the gas phase can effuse from the Knudsen cell trough a small hole. This results in an atomic beam with a large angular spread which is not efficient for applications that require some degree of collimation. This low efficiency leads to an unnecessary high load on the vacuum system and a short lifetime of the source requiring frequent refilling with reactive (alkali-)metals. Placement of a skimming aperture some distance from the source and recycling of the atoms that do not make it trough this aperture can solve these problems. Such recirculating ovens have been constructed \cite{Hau1994,Walkiewicz2000,Pailloux2007} but are reported to be difficult to operate \cite{Bell2010}.

A different, and less complex, approach is the use of a heated tube connected to the Knudsen cell \cite{ScolesBook,Olander1970,Lin2009, Bell2010}. Atoms that hit the wall of the tube are adsorped and, due to the high temperature, emitted again following a cosine angular distribution. Particles entering the tube with a small angle with the axis of the tube as compared to twice the opening angle will hit the wall beyond the middle of the tube and are therefore most probably transmitted. Particles with larger angle however have a larger probability of being sent back. Due to the large angular spread the atoms have when entering the tube, the majority of the atoms will be recycled. Both the recirculating oven and the collimated Knudsen cell leave the center part of the angular distribution, which is most relevant for most experiments, unaltered and thus increase the efficiency of the source.

In ABLIS \cite{Wouters2014} a thermal beam of rubidium atoms is laser-cooled and compressed in the transverse directions to create an atomic beam with a high transverse reduced brightness. When photo-ionized, this beam can be used to create a high-resolution Focused Ion Beam (FIB) surpassing \cite{TenHaaf2014} the current industry standard Liquid Metal Ion Source. At the start of the proposed device a suitable thermal source is required. This source should have a long lifetime and provide an atomic beam that can be laser-cooled and compressed to yield a high-brightness beam with sufficient flux to create a 1 nA ion beam. Furthermore a compact and simple solution is preferred. These requirements lead to the choice of building a collimated Knudsen source. The source presented in this paper is based on the design by Bell et at. \cite{Bell2010}. Some modifications were done to make it more compact and make use of as many off-the-shelf components as possible. Although the design was optimized for our goal it can be used in other applications as well.

In section \ref{sec:Theory} of this paper a summary is given of the relevant theory of the collimated Knudsen cell. Furthermore simulations that are used for comparison with the measurements are explained. The design of the source is described in section \ref{sec:Design}. Section \ref{sec:meas} presents the measurements of the flux, transverse velocity spread and brightness of the source and shows comparisons with analytical and simulated predictions. The conclusions of this work are presented in section \ref{sec:Conclusion}.  

\section{Theory\label{sec:Theory} and simulations}
This section gives a summary of the relevant theory of the collimated Knudsen cell that will later be used for comparison with the experimental results. Furthermore, Monte-Carlo simulations are introduced that will be used for comparison as well.

The flux effusing trough a circular aperture with a radius $r$ from a Knudsen cell in which a gas is present with a density of $n(T)$ is given by \cite{ScolesBook}
\begin{equation}
	\label{eq:FluxTot}
	\Phi_{\text{tot}}={}^1/_4\,n(T)\,\pi r^2\,\langle v_\text{z} \rangle,
\end{equation}
with $\langle v_\text{z} \rangle=\sqrt{8 k_\text{B} T/ \pi m}$ the average speed of a gas in thermal equilibrium in which $T$ is the temperature of the Knudsen cell, $k_\text{B}$ Boltzmann's constant and $m$ the mass of the atom. For alkali metal vapors above the melting temperature, the atomic density can be approximated by \cite{CRC_Vapour}
\begin{equation}
	\label{eq:Density}
	n(T) =p^*\exp(-T^*/T) / \left(k_\text{B} T\right), 
\end{equation}
with $p^*$ and $T^*$ constants for the specific atomic element (the values for rubidium are given in table \ref{tab:Constants}). When a circular collimating tube is attached to the source and the density in the tube is low enough that inter-atomic collisions play no role, the flux out of the tube can be approximated by $\Phi=\Phi_{\text{tot}}\, W$, with the Clausing factor $W$ \cite{Beijerinck1975} given by
\begin{equation}
	\label{eq:Clausing}
	W=\frac{8 r}{3 l+ 8 r}, 	
\end{equation}
in which $l$ is the length of the tube. Although the total flux from the source is reduced, the center-line intensity is unaltered thus increasing the efficiency of usage of the atoms from the source. 

Depending on the density in the Knudsen cell, inter-atomic collisions may play a role on the longitudinal ($l$) or transverse ($d=2\,r$) length scale of the tube. The importance of collisions is determined by the Knudsen number \cite{Olander1970,ScolesBook} which is defined as the ratio between the mean-free-path $\lambda_{mfp}$ and the relevant dimension $x$: 
\begin{equation}
	K_{\text{n},\text{x}}=\frac{\lambda_\text{mfp}}{x} =  \frac{1}{x} \left(4 \sqrt{2} \pi\,{r_\text{vdw}}^2\,n(T) \right)^{-1},
\end{equation}
where $r_\text{vdw}$ is the Van der Waals radius of the atom (given in table \ref{tab:Constants}). The regime in which the collimating tube is operated is described by the Knudsen numbers relating to the length ($K_{\text{n},\text{l}}$) and the diameter($K_{\text{n},\text{d}}$) of the tube:
\begin{itemize}
\item In the transparent regime ($K_{\text{n},\text{d}}>1$ and $K_{\text{n},\text{l}}>10$) no collisions take place between the atoms inside the tube \cite{Olander1970,ScolesBook}. This allows for a description of the angular distribution by geometric arguments and makes simple Monte-Carlo simulations possible \cite{Beijerinck1975}. 
\item In the opaque regime ($K_{\text{n},\text{d}}>1$  and $K_{\text{n},\text{l}}<10$) collisions take place for atoms moving in the longitudinal direction but not for atoms moving across the tube. This effect broadens the angular distribution but does not alter the total flux from the tube \cite{Olander1970,ScolesBook}. 
\item In the continuum regime ($K_{\text{n},\text{d}}<1$ and $K_{\text{n},\text{l}}<10$) collisions also occur between atoms moving across the tube. This will lead to hydrodynamic effects and a reduction in flux \cite{Olander1970,ScolesBook}.
\end{itemize}

\begin{table}
	\small
	\caption{\label{tab:Constants}Atomic constants used in the calculations and the simulations. All data is taken from \cite{SteckRb} except when indicated otherwise. Note that the values for $p^*$ and $T^*$ are only valid for temperatures between 312 K and 550 K. The saturation intensity is averaged over the expected equilibrium population distribution of the m$_{F}$ levels in the F=3 ground state and is valid for linearly polarized light.}
	\begin{tabular}{l|r|r}
		\hline
		Parameter (unit) & Symbol & ${}^{85}$Rb \\
		\hline
		Abundance (\%) & $ab$ & 72.2 \\
		Mass (amu) & $m$ & 84.91 \\
		Pressure constant ($10^9$ Pa) & $p^*$ & $2.05$ \\
		Temperature constant ($10^3$ K) & $T^*$ & $9.30$ \\
		Van-der-Waals radius \cite{CRC_Radii} (pm) & $r_\text{vdw}$ & 303 \\
		Wavelength (nm) & $\lambda$ & 780 \\
		Natural linewidth (MHz) & $\Gamma/2 \pi$ & 6.07 \\
		Saturation intensity (W/m${}^2$) & $I_\text{sat}$ & 31.3 \\
		\hline
	\end{tabular}
\end{table}  

Although analytical expressions have been reported that describe the angular distribution of the atoms \cite{ScolesBook,Olander1970}, the full distribution function is not available in analytical form. Therefore, in order to predict the performance of the collimating tube and to verify the experimental results a Monte-Carlo simulation was set up to find the distribution function of the atoms at the end of the tube $f(x,y,v_\text{x},v_\text{y},v_\text{z})$ for a given geometry ($r$, $l$) and temperature of the Knudsen cell $T_\text{c}$ and collimation tube $T_\text{t}$. These simulations assume a uniform (in $x$ and $y$) circular spatial distribution of atoms at the entrance of the tube with angles chosen to match those of a cosine emitter \cite{ScolesBook}. Then the simulation finds the trajectories of these atoms trough the tube using geometrical calculations while taking the inelastic collisions with the walls (following a cosine emitter perpendicular to the wall surface) into account. In \cite{Wouters2014} this implementation was verified by comparison with theoretical predictions for the angular distribution function. The Monte-Carlo simulations can be used to compare the experimentally determined flux, transverse velocity distributions and brightness to values that can be expected from a source in the transparent regime. 
 
\section{Design\label{sec:Design}}
\begin{figure}[t]
	\includegraphics[width=1.0\linewidth]{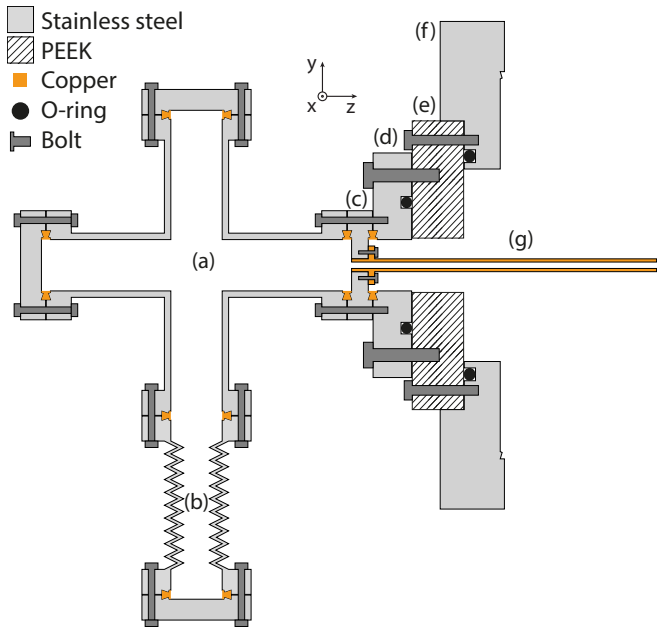}
	\caption{\label{fig:drawing}(Color online) Cross-section of the collimated Knudsen source.}
\end{figure}

Figure \ref{fig:drawing} shows a detailed design of the source, which for easy manufacturing is designed with as many off-the-shelf components as possible: only the tube (g) and thermal insulator (e) needed to be fabricated from scratch while the remaining parts can be used as-is or with minor modifications. The Knudsen cell is formed by several stainless-steel 16 mm Conflat (CF) pieces: a cross piece in the center (a), blank flanges on the top and left and flexible bellows on the bottom (b) in which the rubidium resides. By flexing the bellows a glass ampule filled with solid rubidium can safely be broken while the source is under vacuum. On the right side a double-sided blank 16 mm CF flange (c) has a hole large enough to just fit the copper tube outer diameter (4 mm). The tube (g) is secured to this flange by means of two screws going into the right side of the double-sided flange. The double-sided CF flange is connected to a 16 mm-40 mm CF reducer flange (d) by means of a copper gasket. The other side of the reducer flange is flattened and a groove is made which houses a Viton \cite{Viton} O-ring which is pressed against a thermal insulating plastic ring (e) made out of PEEK (polyether ether ketone) \cite{PEEK}. The usage of this 15 mm thick polymer slab provides a major reduction in the length of the source over bulky ceramic breaks. The insulating ring is connected to a modified 100 mm CF flange (f) sealed again with a Viton O-ring. These O-rings limit the working temperature to 473 K (the PEEK has an upper continuous working temperature of 523 K) which is more than sufficient to create the desired beam of rubidium. The cell is placed inside an insulating box to reduce the heat losses to the surrounding and make the cell less susceptible to influence from outside.  

All four flanges on the cross piece are heated by band heaters \cite{Kurval}. A thermocouple placed on the center of the crosspiece measures the temperature $T_\text{c}$ and a commercial proportional-integral (PI) controller \cite{Omron} keeps it stable within 0.1 K. The flange below the bellows is also heated by a band heater and connected by a thermocouple which allows the temperature $T_\text{b}$ to be stabilised by a second identical PI-controller. To heat the tube a flat tungsten wire is used which is electrically insulated by a silicone heat pad \cite{SilPad} (not shown in the picture). Passing a stable current trough this wire results in a tube temperature $T_\text{t}$, which is measured with a thermocouple attached to the tip of the tube. Vacuum feed-throughs for the heating wires and the thermocouple are placed in the 100 mm CF flange (f).

Having three different temperatures in the system allows for flexibility in operation of the source. Under the assumption that the rubidium collects at the coldest spot in the system the tube should be hotter than the cross and bottom part ($T_\text{t} > T_\text{c}$ and $T_\text{t} > T_\text{b}$) to prevent clogging. At the start of the operation the bottom temperature (where the rubidium resides) can be made higher than the cross ($T_\text{b} > T_\text{c}$) to saturate the walls of the cell and the tube faster. Before opening the cell for maintenance the cross temperature can be made higher than that of the bottom ($T_\text{c} > T_\text{b}$) to ensure most rubidium collects in the bottom of the cell.

\begin{table}
	\small
	\caption{\label{tab:Props}Collimated source dimensions and temperatures}
	\begin{tabular}{l|r|r}
		\hline
		Parameter (unit) & Symbol & Value \\
		\hline
		Tube inner radius (mm) & r & 1.0 \\
		Tube length (mm) & l & 95.0 \\
		Knudsen cell temperature (K) & $T_\text{c}$ & 373 \\
		Tube temperature (K) & $T_\text{t}$ & 393 \\
		\hline
		Theoretical resulting flux ($s^{-1}$) & $\Phi$ & $3.9 \times 10^{13}$ \\
		\hline
	\end{tabular}
\end{table}

\begin{figure}
	\includegraphics[width=1.0\linewidth]{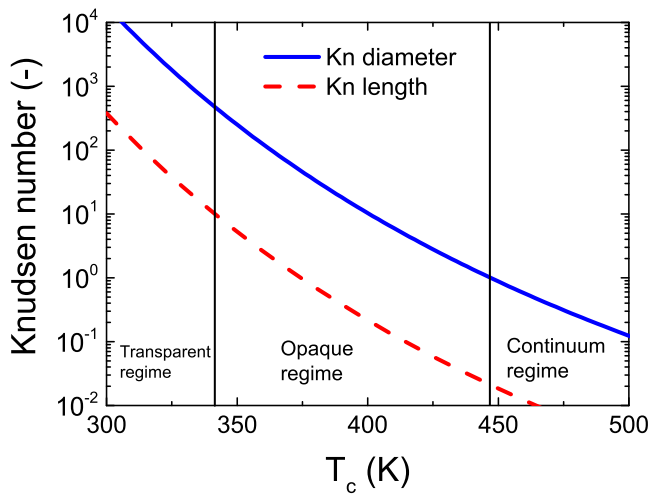}
	\caption{\label{fig:knudsen}(Color online) Knudsen numbers for the tube diameter (blue solid line) and length (red dashed line). In the low temperature part the tube operates in the transparent regime in which collisions do not play a role. In the middle part collisions only play a role at the scale of the tube length and is called the opaque regime. At high temperatures the tube is operated in the continuum regime in which collisions play a role on the length scale of the radius of the tube.}
\end{figure}

The inner dimensions of the collimating tube are based on calculations of the capture range and capture velocity for laser-cooling of rubidium atoms as are described in \cite{Wouters2014}. The capture range of a 5 cm long MOC for rubidium is $r_\text{c}=0.21~\text{mm}$ whereas the acceptance angle, which is based on the capture velocity and average longitudonal velocity, is $\theta_\text{c}=v_c / \langle v_\text{z} \rangle = 7.7~\text{mrad}$. For practical purposes we choose a tube with a radius of 1 mm, which is larger than the capture range. The length of the tube was chosen such that the opening angle $\theta = r/l$ is slightly larger than the acceptance angle of the MOC. These choices lead to a reduction of the capture efficiency of the MOC. Additional simulations indicate that when a cooled and compressed atomic beam with a flux equivalent to 1 nA is desired at the end of the MOC the source should be operated at a temperature of $T_\text{c}$ = 373 K. In this configuration the lifetime of the source is $36 \times$ longer than that of a Knudsen source without collimation. The tube itself should be kept at a temperature $T_\text{t}$ slightly higher than that of the source to prevent clogging. A summary of the parameters of the tube are shown in table \ref{tab:Props}. Figure \ref{fig:knudsen} shows the Knudsen number for the length and diameter of the tube for different source temperatures and indicates the different regimes of operation. At the target temperature of 373 K the tube will be in the opaque regime and thus some broadening of the velocity distribution can be expected. 

\section{Measurements\label{sec:meas}}
The performance of the collimated Knudsen source was evaluated using laser induced fluorescence (LIF). In this experiment a (near)-resonant laser perpendicular to the atomic beam excites the atoms and the resulting fluorescence is imaged onto a CCD camera allowing absolute densities to be measured. By detuning the laser from resonance the transverse velocity profile can be determined by bringing Doppler-shifted atoms back into resonance. Measuring the transverse fluorescence profile as a function of the laser detuning allows the determination of the transverse phase-space density, $\eta(x,v_\text{x})$, which is defined as the amount of flux in a unit of transverse position, $x$, and transverse velocity, $v_\text{x}$:
\begin{equation}
	\eta(x, v_\text{x}) = \frac{d \Phi}{d x\,d v_\text{x}}.
\end{equation}
Integration over $x$ and $v_\text{x}$ yields the total flux of the source $\Phi$, while the position averaged spectral flux density $\psi (v_\text{x}) = d \Phi / d v_\text{x}$ can be calculated by integrating over all transverse positions $x$. The width of this spectral flux density indicates the transverse temperature of the source.

\subsection{Experimental setup}
\begin{figure}
	\includegraphics[width=1.0\linewidth]{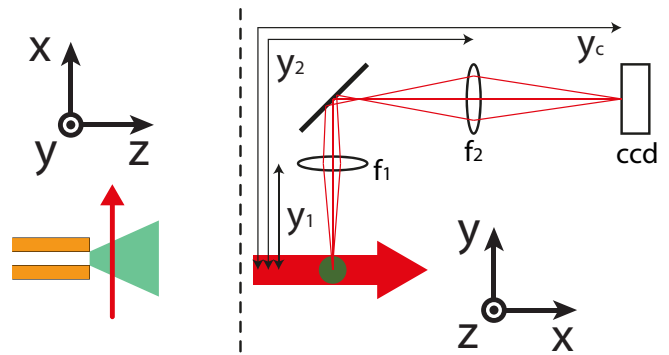}
	\caption{\label{fig:Setup}(Color online) Schematic overview of the experimental LIF setup. At the left a view of the x-z plane is given in which the tube, the diverging atomic beam and the probe laser are visible. On the right the x-y plane with the imaging system is depicted.}
\end{figure}

A schematic overview of the setup used in this experiment is shown in figure \ref{fig:Setup}. A collimated probe beam, resulting from a Coherent 899-21 Ti:Saphire laser, traveling in the x-direction, linearly polarized in the z direction and tuned to the $5\,{}^2P_{1/2}F=3$ to $F'=4$ transition of rubidium-85, is crossed with the atomic beam (traveling in the z-direction) just after the exit of the tube. The imaging system consists of two bi-convex lenses \cite{Thorlabs} which create an image of the emitting atoms from the x-z plane onto a CCD camera \cite{ApogeeU9000} with a magnification of 8.3. The camera captures light from an imaging area 4.00 x 0.14 mm (x times z) and the resulting signal is summed in the z-direction to increase signal to noise ratio. The dimension of the laser beam in the y-direction is larger than the size of the atomic beam so that all atoms can be probed. Using the specified quantum efficiency ($Q_\text{E}$) and gain ($C_\text{epc}$) of the camera, absolute photon count numbers can be determined. In front of the camera a 715 nm longpass filter \cite{ThorlabsFilter} is placed to reduce stray light. Although the imaging system is shielded, some stray light and reflections inside the vacuum vessel reach the camera and contribute to a constant background signal. This is corrected for by taking a measurement at a large detuning (500 MHz), such that no atoms are resonant, and subtracting this from all subsequent measured data.

To gather spectral information, the detuning of the probe laser $\delta=\omega_\text{l} - \omega_\text{a}$, defined as the difference between the laser angular frequency $\omega_\text{l}$ and the atomic transition frequency $\omega_\text{a}$, is scanned from -500 to +500 MHz. The actual frequency of the probe laser is measured by generating a beat note with a second laser that is locked to the atomic transition by means of modulation transfer spectroscopy \cite{McCarron2008}, and measuring the resulting frequency difference with a spectrum analyzer. 

\subsection{Methods}
The following derivation shows how the transverse phase-space density, $\eta(x,v_\text{x})$, can be extracted from the fluorescence. At a given detuning $\delta$ the local flux of photons from the atoms is given by \cite{ScolesBook}: 
\begin{equation}
	\label{eq:SFD1}
	\Phi_{\text{ph}}\left(\delta,x \right)= \langle t_\text{tr} \rangle \, \Gamma \int_{-\infty}^{\infty} \eta(x,v_\text{x}) \, \rho_\text{ee}(\delta, v_\text{x})\,d v_\text{x},
\end{equation}
in which $\Gamma$ is the lifetime of the excited state, $\langle t_\text{tr} \rangle$ is the average transit time of the atoms trough the imaging volume with a longitudinal size of $\Delta z$ and $\rho_\text{ee}(\delta, v_\text{x})$ is the excited state population of atoms with a transverse velocity $v_\text{x}$ and at a laser detuning $\delta$ \cite{FootBook}:
\begin{equation}
	\rho_\text{ee}(\delta, v_\text{x}) = \frac{s_0/2}{1+s_0+4\left(\delta-2 \pi v_\text{x}/\lambda\right)^2/\Gamma^2},
\end{equation}
where $\lambda$ is the laser wavelength and $s_0=I_\text{probe}/I_\text{sat}$ is the local saturation parameter which is constant over the imaging volume. When the velocity distribution is much broader than the power-broadened line-width of the transition $\Gamma\sqrt{1+s_0}$, which is the case for this source and transition, $\eta(x,v_\text{x})$ can be assumed constant over the width of $\rho_{\text{ee}}(\delta,v_\text{x})$ and the phase-space density can be approximated by:
\begin{equation}
	\eta(x,v_\text{x}) \approx \frac{\Phi_{\text{ph}}(\delta,x)}{\Gamma \langle t_\text{tr} \rangle}  / \int_{-\infty}^{\infty} \rho_\text{ee}(\delta,v_\text{x})\,d v_\text{x}.  
\end{equation}
For a beam with $v_\text{z}$ distributed according to a Maxwell-Boltzmann distribution the average transit time $\langle t_\text{tr} \rangle$ is given by $\langle t_\text{tr} \rangle=\langle \Delta z / v_\text{z} \rangle=\Delta z / \sqrt{\pi k_\text{B} T/ (2 m)}$.

 The number of counts on the CCD camera ($C_\text{CCD}$) is given by:
\begin{equation}
	\label{eq:SFD2}
	C_{\text{CCD}}(\delta,x) = \Phi_{\text{ph}}(\delta,x) \, T_{\text{Geom}} \, T_\text{w} \, T_\text{f} \, t_{\text{CCD}} \, Q_\text{E} / C_\text{epc}, 
\end{equation}
in which  $T_{\text{Geom}}=\pi {r_\text{l}}^2 / 4 \pi {y_\text{1}}^2$ is the part of the isotropic emission sphere that the first lens covers. The other experimental parameters are explained and given in table \ref{tab:Exp}. Combining equations \ref{eq:SFD1} and \ref{eq:SFD2} yields the phase-space density of ${}^{85}$Rb atoms that were in the F=3 ground state. This number needs to be corrected for the initial thermal F=2 and F=3 ground state distribution in rubidium-85 (factor 12/7) and its abundance (factor $1/ab$) to get the total number of rubidium atoms.

With the phase-space density $\eta(x,v_\text{x})$ known, the emittance in the x-direction $\epsilon_\text{x}$ can be calculated using equation 10 from ref \cite{Luiten2007}. Assuming that the atomic beam is ionized into a beam with current $I=e \Phi$ and that the emittance in the y direction equals that of the x direction the transverse reduced brightness of the atomic beam can be calculated by using equation 12 from ref \cite{Luiten2007}. 

Although all experimental parameters were measured with care, calculating absolute densities from LIF experiments should be performed with caution: several assumptions that are taken are not fully correct; for example the value for the saturation intensity $I_\text{sat}$ which assumes the atoms are pumped into a equilibrium $m_F$-distribution. Furthermore the assumption is made that emission is isotropic and that all atoms are in focus of the imaging system.

Measurements were performed for several sets of temperatures for the different components: the cross and bottom temperature are set equal ($T_\text{c}=T_\text{b}=T$) and the tube temperature was made 20 K higher than that of the cross and bottom ($T_\text{t}=T+20\,\text{K}$). In the remainder of this article we refer to the temperature of the cross and bottom as 'the temperature'. After changing the temperature of the source at least one hour of delay while monitoring the flux was taken to make sure the flux was stable. For each temperature the shutter time of the camera was altered such that a high signal to noise ratio was achieved without over-exposure.

\begin{table}
	\small
	\caption{\label{tab:Exp}Experimental parameters. All optical properties are given at a wavelength of 780 nm.}
	\begin{tabular}{l|r|r}
		\hline
		Parameter (unit) & Symbol & Value \\
		\hline
		Image z-position after tube (mm) & $z_\text{i}$ & 2.5 \\
		Probe intensity (W/m$^2$) & $I_\text{probe}$ & 44 \\
		Image z-slice width (mm) & $\Delta z$ & 0.14 \\
		Vacuum window transmission (\%) & $T_\text{w}$ & 92 \\
		Filter transmission \cite{ThorlabsFilter} (\%) & $T_\text{f}$ & 83 \\
		First lens y-position (mm) & $y_\text{1}$ & 214 \\
		First lens focal length (mm) & $f_\text{1}$ & 126 \\
		First \& second lens radius (mm) & $r_\text{l}$ & 2.54 \\
		Second lens y-position (mm) & $y_\text{2}$ & 584 \\
		Second lens focal length (mm) & $f_\text{2}$ & 60 \\ 
		Camera y-position (mm) & $y_\text{c}$ & 997 \\
		Camera electrons per count \cite{ApogeeU9000} (-) & $C_\text{epc}$ & 1.5 \\	
		Camera quantum efficiency \cite{ApogeeU9000} (\%) & $Q_\text{E}$ & 50 \\
		Camera shutter time (s) & $t_{\text{CCD}}$ & 0.15 - 2.50\\
		\hline
	\end{tabular}
\end{table}  

\subsection{Results}
\begin{figure}
	\includegraphics[width=1.0\linewidth]{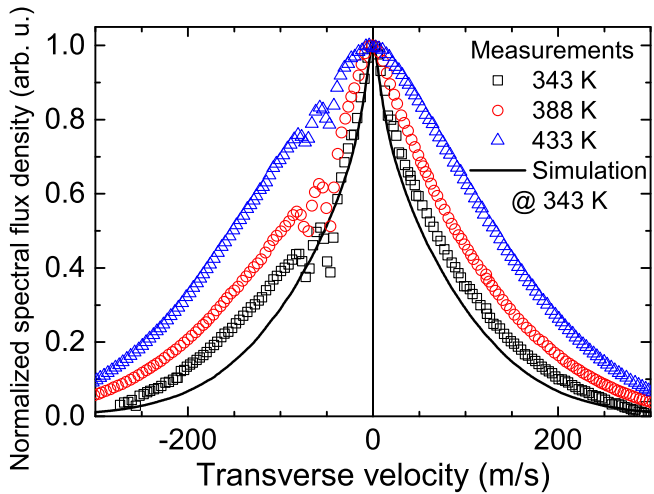}
	\caption{\label{fig:spectrum}(Color online) Measured (scatter) and simulated (line) spectral flux density for different source temperatures. The measured and simulated data for $T_\text{c}$ = 343 K show good agreement. As temperature increases the velocity distribution is broadened.}
\end{figure}

The spectral flux density was determined for different temperatures by measuring the phase-space density and integrating over $x$. Figure \ref{fig:spectrum} shows the resulting spectral flux density for different temperatures as well as a Monte-Carlo simulation at a temperature of 343 K. The first observation that can be made is the asymmetry in the measured data: the wing of the negative part of the profile seems broader than the positive side and two dips appear there as well. The broader wing can be explained by additional fluorescence from atoms resonant with the weaker F=3 to F'=3 and F=3 to F'=2 transitions. The dips, located at the crossover transitions \cite{FootBook} F=3 to F'=3,4 at -60 MHz, F=3 to F'=2,4 at -92 MHz, are assumed to be due to saturation effects from part of the probe beam that is reflected on the uncoated vacuum window.

\begin{figure}
	\includegraphics[width=1.0\linewidth]{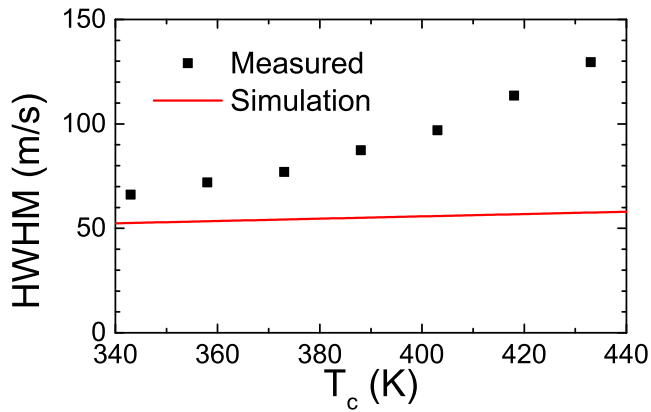}
	\caption{\label{fig:width}(Color online) Half-width-half-maximum (HWHM) width of the transverse velocity distribution for different source temperatures as measured (black filled squares) compared to simulations (red solid line). Increasing the temperature results in a broader distribution.}
\end{figure}

The right side of the measurement at 343 K does resemble the simulated data indicating that the source is indeed operated in the transparent regime. For higher temperatures the velocity distribution is broadened which is due to the operation of the tube in the opaque regime (see figure \ref{fig:knudsen} for the Knudsen numbers at which the experiment was performed) which results in broadening due to collisions in the tube. The broadening was characterized by calculating the half-width-half-maximum (HWHM) of the positive side of the distribution due to the non-ideal effects (dips and broadening) at the negative side. Figure \ref{fig:width} shows these HWHM numbers for different operating temperatures of the source. Here the effect of collisions within the tube can be seen when comparing to the collisionless simulation data: the higher the source temperature, the higher the pressure in the tube giving rise to more collisions which broaden the velocity distribution.  

\begin{figure}
	\includegraphics[width=1.0\linewidth]{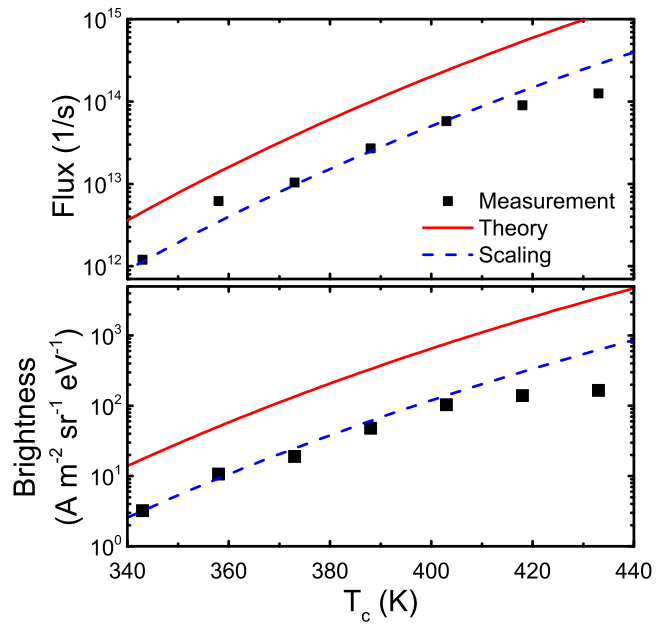}
	\caption{\label{fig:FluxBr}(Color online) Measured total (abundance and F-state corrected) rubidium flux and 10\%-brightness (black filled squares) compared to the theoretical prediction for a tube in the transparent regime (red solid line). Both the flux and brightness are somewhat lower than predicted but for temperatures upto $T_\text{c}$ = 420 K both quanteties do scale according to the predictions as is indicated by the blue dashed lines.}
\end{figure}

Integration of the spectral flux density results in the total flux from the collimated Knudsen source. The upper part of figure \ref{fig:FluxBr} shows the flux for different temperatures compared to the theoretical prediction of equations \ref{eq:FluxTot}-\ref{eq:Clausing}. As can be seen from the blue dashed line the scaling of the flux with the temperature agrees well with the predictions. From 420 K the measured flux was somewhat lower than the scaling law which can be explained by looking at the Knudsen numbers, which are plotted for different temperatures in figure \ref{fig:knudsen}. At the highest temperature (440 K) the Knudsen number related to the diameter of the tube equals 2 indicating that continuum flow regime ($K_{\text{n},\text{d}}<1$), in which the flux is lowered, is approached. Although the scaling agrees well with the predictions, the absolute flux was found 4 times lower than expected which may partly be justified by the simplification of the physics of the LIF experiment. A lower flux can also be explained by a cold spot in the Knudsen source. However, to justify the $4\,\times$ difference in flux the temperature at this cold spot should be 15 K lower than the measurement from the thermocouple which seems improbable due to the use of thermally conductive materials. 

\begin{figure}
	\includegraphics[width=1.0\linewidth]{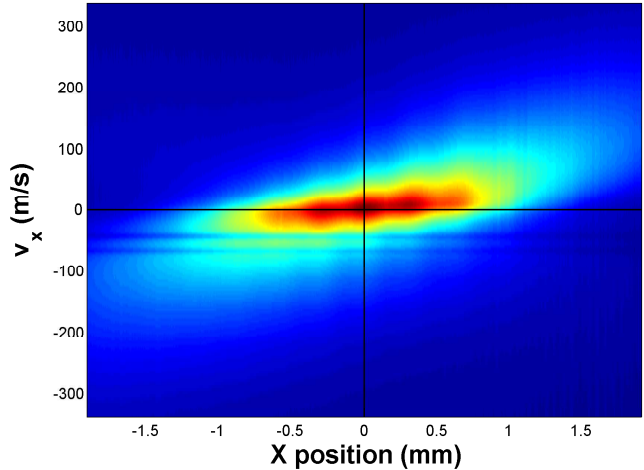}
	\caption{\label{fig:phasespace}(Color online) Phase space image for the source at  $T_\text{c}$ = 373 K. Propagation of the atoms from the source to the imaging area leads to shearing of the phase space image.}
\end{figure}

In figure \ref{fig:phasespace} a phase-space image of the source at a temperature of 373 K is shown. The horizontal lines at negative velocities are caused by the crossover transitions as before. The phase-space distribution is sheared due to the propagation from the end of the tube to the imaging area over distance $z_i$ from table \ref{tab:Exp}. From the phase-space image the transverse reduced brightness for an ion beam with the same properties as the measured atomic beam was calculated for the 10\% fraction of particles closest to the x-axis for this gives a good indication of the brightness of the center part of the beam. The lower part of figure \ref{fig:FluxBr} shows the 10\%-brightness for different temperatures of the source, which shows the same scaling with temperature as the simulated data upto 420 K. The major part of the discrepancy in the absolute value of the brightness can be justfied by the $4\,\times$ lower than expected flux that was measured. The remaining missing factor 1.7 may be explained by the broader velocity distribution due to collisions in the tube that are not included in the simulations. Still, with a maximal achievable initial brightness of $1.7 \times 10^2$\,A/(m${}^2$\,sr\,eV) at a temperature of 433 K and a brightness increase due to laser cooling of $10^5\times$ \cite{Wouters2014}, an atomic beam brightness of $10^7$\,A/(m${}^2$\,sr\,eV) should be possible.

\section{Conclusion\label{sec:Conclusion}}
In this paper a design of a compact collimated Knudsen source was discussed. This source will be used to load the magneto optical compressor of the atomic beam laser cooled ion source \cite{Wouters2014,TenHaaf2014}. The flux, transverse velocity spread and the brightness were measured using laser-induced fluorescence. The scaling of the flux and brightness with the temperature showed good agreement with the simulated data. The absolute values of the flux and brightness were however lower than expected. This may partly be explained by collisions in the collimation tube that are not taken into account in the simulations. The source is able to produce the desired flux of $4\times10^{13}$~s$^{-1}$ and a transverse reduced brightness of $1.0 \times 10^2$\,A/(m${}^2$\,sr\,eV) at a temperature of 403 K which is sufficient to reach our goal \cite{Wouters2014}.

\begin{acknowledgments}
This research is supported by the Dutch Technologiestichting STW, applied science division of the ``Nederlandse Organisatie voor Wetenschappelijk Onderzoek (NWO)'', FEI Company, Pulsar Physics and Coherent Inc. 
\end{acknowledgments}

\bibliography{Source}

\end{document}